\documentclass[pra,twocolumn,showpacs,superscriptaddress,preprintnumbers,amsmath,amssymb,floatfix]{revtex4}

\usepackage{graphicx}
\usepackage{dcolumn}
\usepackage{bm}

\begin{document}

\newcommand\bsj{\left( \begin{array}{rrr} }
\newcommand\esj{\end{array} \right)}



\title{A scalable, high-speed measurement-based quantum computer using trapped ions}
\author{Ren\'e Stock}
\email{restock@physics.utoronto.ca}
\affiliation{{Department of Physics, University of Toronto, Toronto, Ontario M5S 1A7, Canada}}
\author{Daniel F. V. James}
\affiliation{{Department of Physics, University of Toronto, Toronto, Ontario M5S 1A7, Canada}}

\date{\today}


\begin{abstract}
We describe a scalable, high-speed, and robust architecture for measurement-based quantum-computing with trapped ions. Measurement-based architectures offer a way to speed-up operation of a quantum computer significantly by parallelizing the slow entangling operations and transferring the speed requirement to fast measurement of qubits. We show that a 3D cluster state suitable for fault-tolerant measurement-based quantum computing can be implemented on a 2D array of ion traps. We propose the projective measurement of ions via multi-photon photoionization for nanosecond operation and discuss the viability of such a scheme for Ca ions. 
\end{abstract}

\maketitle


The rapid progress in quantum information processing systems has been fueled by the realization that the algorithmic complexity of a quantum computer scales polynomially with the size of certain important problems rather than exponentially~\cite{Nielsen:Chuang}. This presents a tremendous advantage for large problems that are so far solvable only on a time scale of years. However, the practical utility of large-scale quantum computers will also depend on their ability to compete with current classical computers.  Consider, for example, Shor's factoring algorithm - the composite integer RSA640 ($N=640$ bits) requires $\approx32\; N^3=8.4\;10^9$~\cite{Fowler:2004} operations (neglecting error correction overheads and improved scaling with other trade-offs~\cite{Shor:improve}). To compete with a distributed network that can factor RSA640 in 5 months~\cite{RSA}, quantum operations on time scales of 1.5~ms are required; to factor RSA640 in, say, 5 minutes, operation time scales have to be improved to 36~ns. Error correction overheads will further worsen time scales. In other words,  {\it nanosecond operations are essential for practical large-scale quantum computers}.

In this Letter, we address this speed issue for one of the most promising quantum computing (QC) implementations, namely, ion-traps~\cite{Cirac:1995}. In contrast to standard quantum circuit schemes considered so far for ion-traps, we consider measurement-based QC paradigms~\cite{Raussendorf:2001}, where the actual processor speed is mostly determined by the measurement time scales. We demonstrate (i) this one-way quantum computing (1WQC) scheme has significant advantages for ion-trap QC; (ii) that a 3D cluster state for fault-tolerant computing can be efficiently implemented in 2D ion-trap architectures; and (iii) that multi-photon ionization and detection of the emitted electron using the ion trap potential as a guide can significantly speed up high-efficiency measurements  to nanosecond time scales.

Ion-trap QC~\cite{Cirac:1995} has surpassed several major milestones on the QC roadmap~\cite{roadmap}. Recent experiments have entangled up to eight ions~\cite{Haffner:2005}, demonstrated above $99\%$ fidelity gates~\cite{Benhelm:2008,SchmidtKaler:2003}, and coherence times of 10-34~s \cite{Langer:2005,Haffner:2005:ApplPhysB}. Distant entanglement of ions via interference of emitted photon pairs has been demonstrated~\cite{Moehring:2007} and scalable chip-based trap architectures are being implemented~\cite{Kielpinski:2002,iontraps}. However, time scales for logic-gate operations are slow, on the order of 1-100$\mu s$ for entangling gates~\cite{Cirac:1995,fastgates:all} and $1 \mu s$ to 10ms for single-qubit operations and measurement~\cite{SchmidtKaler:2003}. Moreover, the shuttling of ions required in scalable architectures puts even worse timing constraints on two-qubit gates (50-100$\mu s$)~\cite{Hucul:2008}. Measurement-based QC paradigms~\cite{Raussendorf:2001} offer a way around this because the computational resource, a multipartite entangled cluster state, can be created via entangling operation applied in parallel and offline. This has the tremendous advantage that the usual requirement to avoid motional heating in ion traps is restricted to this first entangling step and is removed from the actual QC process. In the 1WQC protocol, processing of information occurs through measuring qubits in a prescribed  basis, combined with feedforward of measurement outcomes. The processor speed is determined by measurement, readout, and classical processing time scales which, as we show, should be possible on a nanosecond time scale. Methods for error correction have recently been proposed which introduce the necessity for 3D cluster states~\cite{Raussendorf:2006}. We describe how a 3D cluster can be efficiently implemented in a 2D architecture {\it without} resorting to time as third dimension~\cite{Raussendorf:2007}.


In the 1WQC paradigm, all entanglement operations are done in parallel and offline before commencement of the algorithm. This multipartite entangled cluster state is created by shuttling ions in a 2D lattice geometry into close proximity and by applying standard controlled-PHASE (CPHASE) gates~\cite{Cirac:1995,fastgates:all,Milburn:2000} between neighboring atoms.  In such a 2D array, measurements of the qubits in different bases and feedforward of measurement outcomes allow the simulation of a universal quantum circuit~\cite{Raussendorf:2001}. While this 2D cluster state represents a universal resource, topological encodings in 3D structures are required for error correction with high thresholds~\cite{Raussendorf:2006}. More recently, quasi-2D fault tolerant cluster states, where the third dimension is replaced by time, were introduced~\cite{Raussendorf:2007}, effectively mapping a 3D cluster to a 2D array. However, in this scheme, the two-qubit gates required during the measurement-based QC will nullify the potential for speed-up over circuit based QC. 

\begin{figure}[h]
  \centering
  \includegraphics[width=85mm]{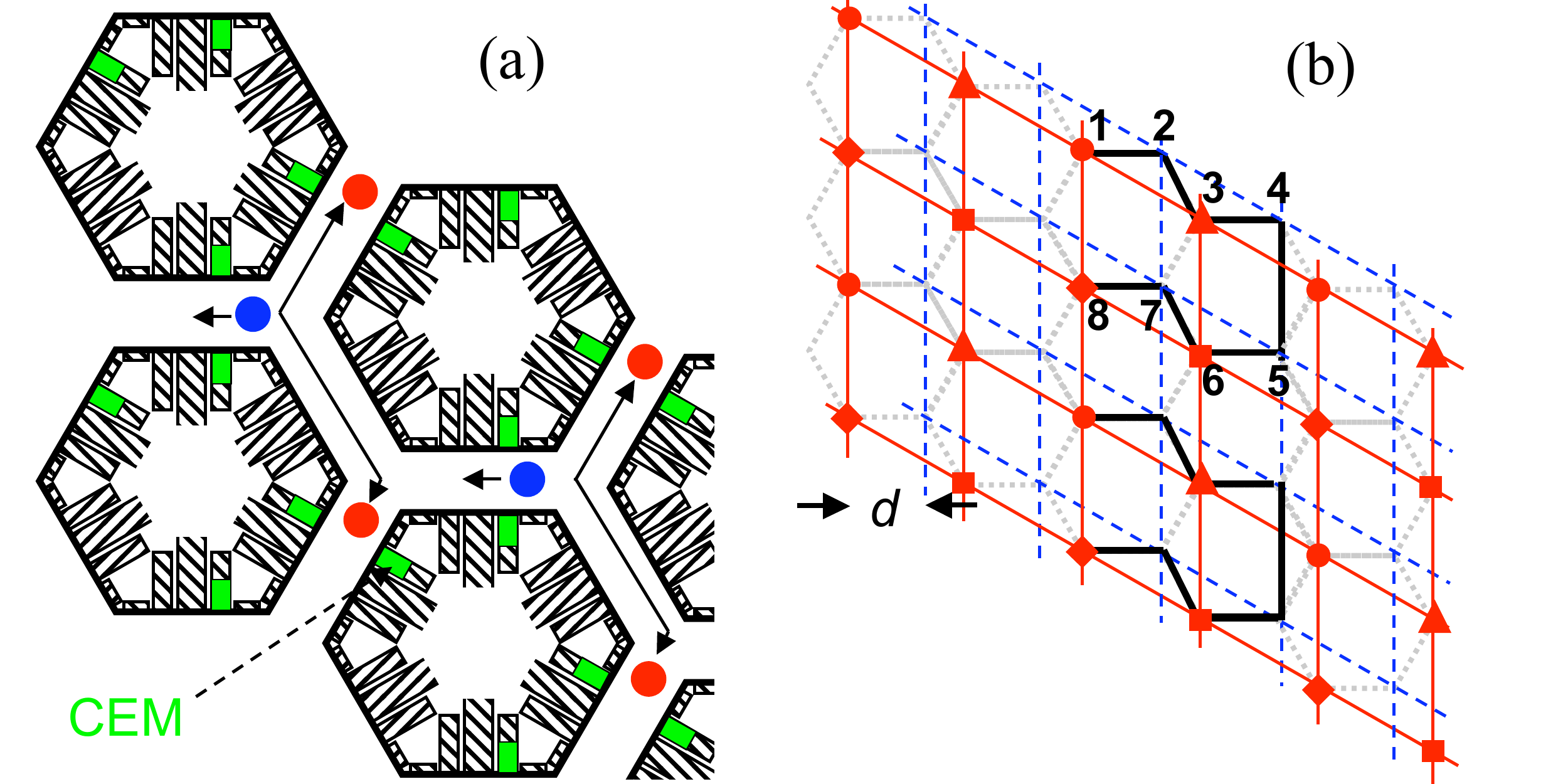}
  \caption{(a) Hexagonal ion trap architecture with $120^\circ$ junctions. Single channel electron multiplier (CEM) detectors for high-efficiency electron detection are indicated. For entangling operations, atoms can be shuttled into close proximity with high fidelity to remain in the ground state. To entangle non-nearest neighbor atoms, one atom could be moved out of the way as indicated by arrows. Alternatively, swap operations between qubits at a junction could minimize movement of atoms. (b) Underlying hexagonal array and resulting cluster states for non-nearest neighbor interactions. The hexagonal array is decomposed into two rhombic lattices (red solid and blue dashed lines show the edges of the cluster state layers). Each lattice can again be broken down into further sublattices. For example, 4 sublattices of the solid line rhombic lattice are indicated by circles, squares, triangles and diamond. These, together with 4 sublattices of the dashed rhombic lattice can create an eight-layer 3D cluster state (see text for details). Qubits belonging to the eight different layers are numbered from 1 to 8. (Color online.)}
\label{fig1}
\end{figure}

Because of the requirements of laser access, implementing a full 3D architecture for fault-tolerant 1WQC using ion traps would be extremely challenging. Instead, a 3D cluster state can be implemented efficiently in a 2D ion-trap array with non-nearest neighbor entanglement operations. The architecture required for implementing the 1WQC paradigm is similar to the scalable ion-trap architectures proposed in~\cite{Kielpinski:2002}. This architecture requires inherently slow shuttling of ions~\cite{Hucul:2008}. For faster transport, better fidelity, and to keep ions in the motional ground state, 120$^\circ$ Y-junctions in a 2D hexagonal array (Fig.\ref{fig1}) are preferable to 90$^\circ$ junctions in a square lattice~\cite{iontraps}. 

A full 3D cluster state can be created in this universal \cite{VandenNest:2007} hexagonal 2D architecture in the following way: If we do not restrict ourselves to nearest neighbor entangling operation, then the hexagonal lattice can be broken down into two rhombic lattice sublattices with distance 2d between sites as illustrated in Fig.\ref{fig1}(b). Each of these two sublattices then represents a single 2D layer of a 3D cluster state. To create each layer of cluster state, atoms in each sublattice have to be entangled with its sublattice neighbors, $2d$ away, as described in Fig.\ref{fig1}(a). In each sublattice, a sequence of four CPHASE gates applied in parallel across the array entangles every ion with all four sublattice neighbors, creating a full 2D cluster state layer~\cite{Raussendorf:2001}. Entanglement between different layers can be accomplished by simply entangling ions belonging to different sublattices (indicated by numbers 1-8) in just two parallel applications of a series of CPHASE gates as indicated by thick black lines in Fig.1(b)]. The number of layers can be increased from 2 to $2 n^2$ by increasing the elementary cell of each sublattice by a factor $n$. This corresponds to entangling atoms separated by a distance $2nd$ [see Fig.1(a)]. Note that the distance only scales as $2n$ whereas the number of layers is $2n^2$. An example of an 8-layer system is shown in Fig.1(b) for a distance of $4d$ between entangled atoms. Another important feature is the possibility to entangle the first and last layer of the 3D cluster state, thereby directly and efficiently creating a topological structure as required in~\cite{Raussendorf:2006,Raussendorf:2007}. The number of parallel operation and thus the time to create this 3D cluster state offline is constant in the number of qubits, and only requires six CPHASE gates cycles applied in parallel to all neighboring qubits. 

In addition to standard entangling operation between nearby ions, probabilistic distant entangling operations ~\cite{Moehring:2007,Barrett:2005} could be used to connect 3D ``sub''-clusters, which are created in separate locations and contain $10^3$-$10^4$ qubits. Building up a large cluster this way 1) avoids correlation errors across the entire cluster state, 2) protects the unmeasured cluster from disturbance due to the ionization measurements, and 3) allows for slightly delayed preparation to avoid storage errors as well as re-preparation of subclusters while others are measured. This fulfills the requirements of the more stringent error model in~\cite{Raussendorf:2006}, which includes errors due to correlations and storage (error model 2~\cite{Raussendorf:2006}). Due to disparate time scales between measurement ($\sim$ns) and decoherence time ($\sim$s) storage errors are kept well below $10^{-4}$ even when measuring up to $10^4$ qubits. In addition, decoherence free subspaces have been successfully demonstrated for ions~\cite{Haffner:2005:ApplPhysB} as well as for 1WQC~\cite{Prevedel:2007}. A detailed error analysis with accurate thresholds and overheads is planned for future work.

Measurements of atomic qubits are usually achieved in two steps: A single-qubit rotation, followed by projection into one of the two-qubit states. We assume encoding in the optical $S$-$D$ transition in Alkali-metal-like ions to facilitate nanosecond single qubit rotations and readout using light pulses with a bandwidth much smaller than the energy separation between the $S$ and $D$ states. To increase decoherence time a mixture of encodings would be conceivable, i.e. long-time storage in the hyperfine state and short-time storage in $S$-$D$ states for faster manipulation. Transitions between the $S$ ground state and metastable $D$ state are electric-dipole forbidden with long coherence times ($\sim$s). Single-qubit rotations on the $S$-$D$ transition can be achieved via quadrupole-allowed transitions or two-photon transitions using the intermediate $P_{1/2}$ or $P_{3/2}$ states. Quadrupole Rabi frequencies $\Omega^{E2}_{SD}$ in current experiments are 35.5 kHz for fairly low irradiance of about $6$W/cm$^2$~\cite{SchmidtKaler:2003}. To achieve manipulations on a 2~ns time scale via quadrupole fields would require extremely high irradiances of $10^9\text{W/cm}^2$. For Raman transitions, detuned $10^4$ line widths from the $P$ states, an irradiance of $10^5\text{W/cm}^2$ is sufficient to achieve manipulations on a 1~ns time scale.

\begin{figure}[h]
  \centering
  \includegraphics[width=80mm]{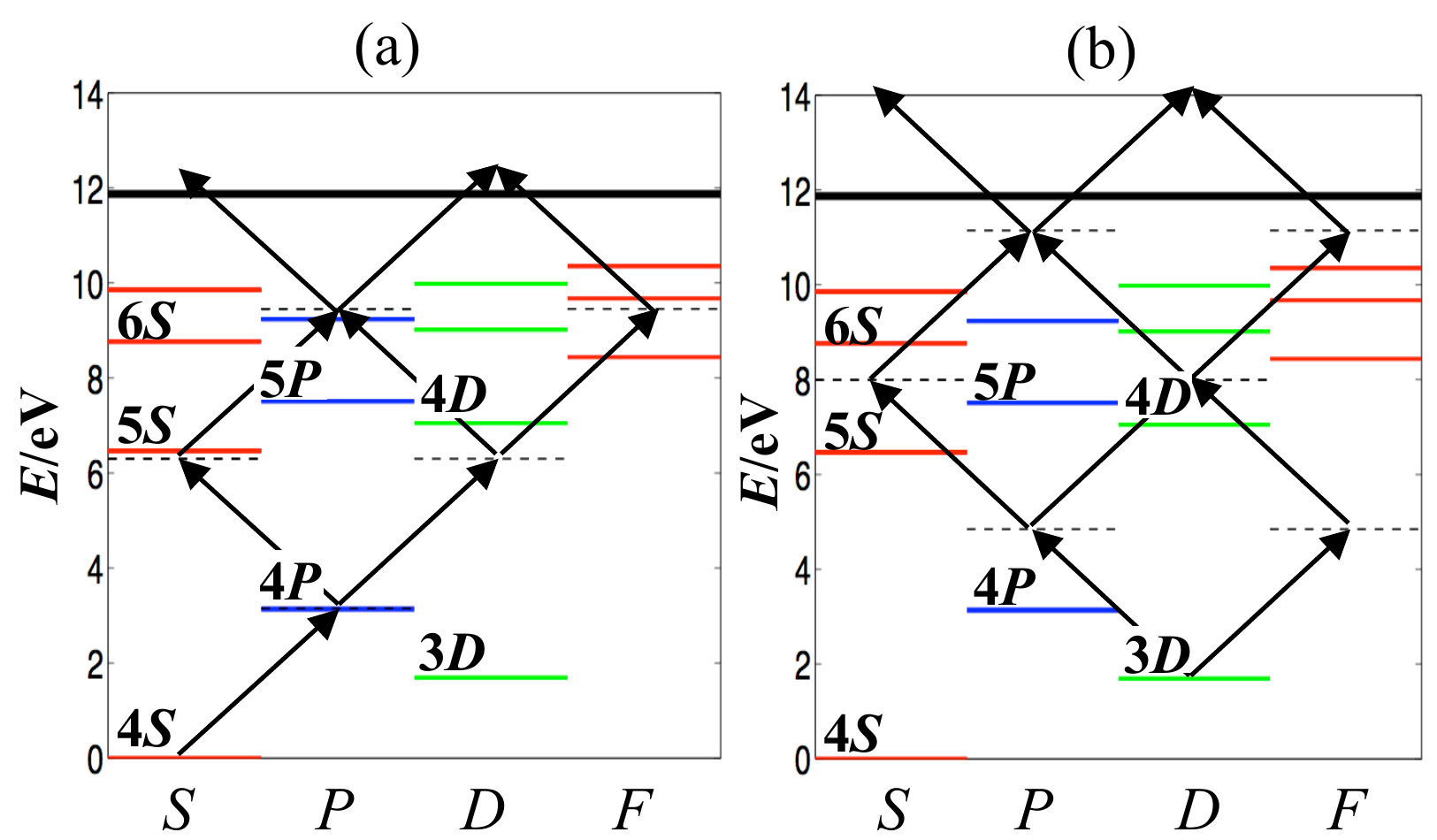}
  \caption{Relevant levels for ionization-readout scheme of the (a) $S$-state and (b) $D$-state in $^{43}$Ca$^+$. Contributing transitions are shown with black arrows. The ionization threshold at 11.87 eV is indicated as thick solid line. (Color online.)}
\label{fig2}
\end{figure}

A more severe timing limit and one that requires a more drastic shift in approach, is due to the fluorescence readout in ion systems where one of the two qubit states is coupled to a shelving state via a cycling transition~\cite{SchmidtKaler:2003}. The time scales for this are limited by the lifetimes of dipole transitions, and photon-collection and photodetection efficiencies to a few microseconds. However, state-dependent multiphoton ionization and subsequent detection of the electron allow measurement on nanosecond time scales. Consider a resonant four-photon ionization of Ca$^+$ in the $S$-state (see Fig.~\ref{fig2}) using easily accessible $380-410$nm excitation wavelengths. From the level diagram in Fig.~\ref{fig2}, it is immediately apparent that a multiphoton ionization transition from the $4S_{1/2}$-ground state should be possible for fairly low intensities. On the other hand photoionization of the $D$-level is strongly suppressed, as the detunings are an order of magnitude larger for each transition. For $380-410$nm, the four-photon transition is close to resonance with several levels for ionizing atoms in the $4S_{1/2}$ state. We can choose a resonance condition with the $4P_{1/2}$ state at 397nm, the $5S_{1/2}$ state at 383nm, or the $6P_{1/2}$ and  $6P_{3/2}$ states at  403nm. A broadband frequency-doubled Ti:Sapphire with appropriate pulse shaping would be able to address all three resonances and exploit interference between ionization paths to improve the ionization fidelity and state discrimination~\cite{Papastathopoulos:2005}. 

Using the effective operator method~\cite{Crance:1978}, the resonant ionization probability for the $S$-state when applying laser light of irradiance $I$ and $\pi$ polarization is $N_\pi^S=\sum_{\lambda}4 \pi  I^4 (J_\pi^\lambda)^2 + 4 \pi I^2 K^2/L^2 $. The nonresonant contributions are described by transition probabilities $J_\pi^\lambda$ for transitions to continua states with different angular momentum $\lambda=S,\;P,\;D,$ etc.; $L$, $K$ are resonant transition operators~\cite{Crance:1978}. For the nonresonant $D$-state ionization, only an incoherent sum of the transition rates, $N_\pi^D=\sum_{\lambda}4 \pi (J_\pi^\lambda)^2 I^4$ need be considered. We include only near-resonance transitions in our estimate of the ionization rates~\cite{NIST} (non-listed matrix elements have been estimated by simple scaling).  Using this basic estimate, we expect the nonresonant parts of photoionization probability to be a factor of at least $[(J_\pi^\lambda)_S/(J_\pi^\lambda)_D]^2=1600$ lower for atoms in the $D$ state compared to the $S$ state. This difference allows a very high fidelity state discrimination as would be necessary for a projective measurement. At resonance with the  $6P_{1/2}$ state, we expect ionization rates on the order of $N_\pi^S\approx10^{9}-10^{10}$s$^{-1}$ for peak irradiances of about $I=10^9$W/cm$^2$.  This allows for ionization on a femto- or picosecond time scale with accessible laser intensities~\cite{Papastathopoulos:2005}.


\begin{figure}[h]
  \centering
  \includegraphics[width=85mm]{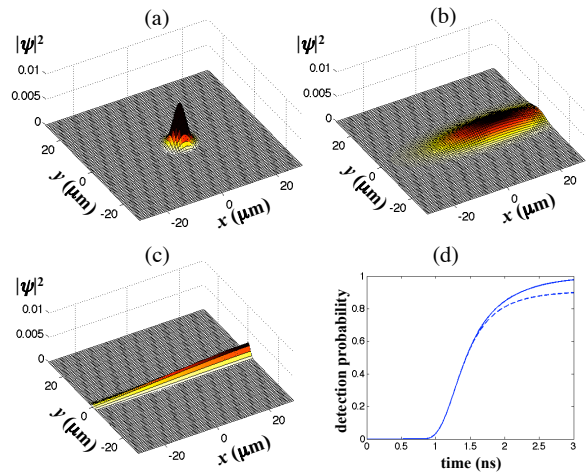}
  \caption{(a-c) Time evolution of the emitted electron wavepacket in the ion-trap saddle potential at three different times, (a) 0.25~ns (b) 1.0~ns (c) 1.5~ns. We assume an initial 4-photon kick of the electrons in the positive x-direction. (d) If we place one detector $30 \mu$m away on the x-axis, we can detect the electron with up to $90\%$ fidelity [see dashed line in (d)]. A second detector can be placed on the negative x-axis to detect the part of the wave packet that escaped along the saddle in this direction. The detection efficiency is then above $99\%$ within a 3~ns time window [solid line in (d)]. (Color online.)}
\label{fig3}
\end{figure}

Photoionization and detection of ions has been used in multiple experiments, e.g. for counting of atoms~\cite{Campey:2006}, imaging of a gas of atoms~\cite{Robert:2001} and even single atoms in a cloud~\cite{Gericke:2008}. Detection of electrons via a single channel electron multiplier (CEM) is generally more efficient ($>99\%$ efficiency) than detection of the heavy ion~\cite{Seah:1990}. A CEM also features detector dead times of nanoseconds or less~\cite{Seah:1990}. Not only is the detection probability higher for the freed electron compared to the ion, but electrons move faster, so that the time between ionization and detection is less. Stability of a particle in the ion trap requires usually the characteristic trap parameter $q_x$ to be in the range $0<q_x<1$. The electron is unstable in the dynamic trapping potential as $q_x^e\propto \sqrt{e/m_e}$ is two orders of magnitude larger than the one for the ion $q_x^\text{ion}\propto \sqrt{e/m_\text{ion}}$. The rf frequency $\omega_\text{rf}$ for the ion-trap field is typically on the order of 20-30 MHz~\cite{SchmidtKaler:2003} so that the characteristic time scale for the electron motion $t_0\approx 1/\omega_\text{rf} \sqrt{m_e/m_\mathrm{ion}}$ in the ion trap is about 0.5~ns. The ion-trap field is therefore time-independent on the electron detection time scale, i.e., the electron moves in an almost time-independent saddle-point potential and will be ejected along the saddle. To describe the motion of the electron in this potential, assume all momentum from the four photon ionization is transferred to the electron in the positive x-direction (the inertia of the {\em trapped} ion is effectively infinite); the corresponding initial electron velocity is then $7\times10^3$m/s. The electron is ejected when the dynamic trap potential is anti-trapping in the x-direction and trapping in the y-direction (with a static weak trap field in the longitudinal z-direction~\cite{SchmidtKaler:2003}). A time-dependent simulation of the electron wavepacket dynamics shows that the ion-trap potential can act as a guide for the electron onto the CEM detectors (see Fig.3). Our simulations show that the electron remains localized in the y-direction with the typical wave packet breathing due to squeezing in the y-direction. In the x-direction, the electron wave packet spreads and ``slides''  down the saddle of the trapping potential due to the initial momentum kick of the photoionization photons. If we place two $20\mu$m size CEM detector about $30\mu$m away from the trap center close to the trap-electrodes on the positive and negative x-axis, we can estimate above $99\%$ detection efficiency within 3~ns [Fig.\ref{fig3}(d)]. Due to trap geometry, placement of detectors may be restricted, but the initial conditions (direction of photoionization beam, timing of ionization) should make it possible to guide the electron into the detector using the existing trap-field. 


For above ionization readout, we also need to consider what happens to the remaining doubly ionized Ca$^{2+}$ atom. The trap-parameters $q_x$ which changes by a factor of $\sqrt{2}$ from Ca$^+$ to Ca$^{2+}$ can be chosen so that  both Ca$^+$ and Ca$^{2+}$ are trapped in a stable regime~\cite{SchmidtKaler:2003}. However, practical constraints such minimization of micromotion~\cite{SchmidtKaler:2003}, may dictate parameters that would result in the loss of Ca$^{2+}$. The relevant time scale for the ion to become unstable in the trap is slower than $1/\omega_\text{rf}$ (0.1-1$\mu$s). The measurement time is much shorter and qubit readout for the one-way QC process will have moved on to other qubits, 10-100 qubits farther away, so that measurements are not affected. Finally, each ion is located in separate regions of the architecture (several $100\mu$m apart) and coupling to other ions can be neglected. The one draw back of this scheme is that the trap would have to be reloaded before running another algorithm. Distant entanglement~\cite{Moehring:2007} can be used to connect smaller clusters which could be reloaded after usage while others are measured.


Our results certainly indicate that measurement-based quantum computing with nanosecond measurement times presents a extremely promising solution to the speed constraints in quantum computing. The proposed nanosecond readout is immediately applicable to loophole-free tests of Bell-inequalities using ions only few meters apart~\cite{loophole:all}. For quantum computing, a detailed experimental analysis is needed to determine realistic error rates, and more theoretical work is on it way to determine accurate error correction thresholds and overheads.


We thank R. Raussendorf, J. Harrington, J. Chiaverini, R. Van Meter, and S. Ghose for helpful discussions. We also thank the referees for their suggestions and their critique. This work is supported by NSERC and the U.S. Army Research Office, Award No. W911NF-05-1-0397. 


\end{document}